\documentclass[journal]{IEEEtran}
\ifCLASSINFOpdf
\else
\fi
\usepackage{gensymb}
\usepackage{graphicx}

\usepackage{amsmath}
\usepackage{amssymb}
\ifCLASSOPTIONcompsoc
  \usepackage[caption=false,font=normalsize,labelfont=sf,textfont=sf]{subfig}
\else
  \usepackage[caption=false,font=footnotesize]{subfig}
\fi
\begin{document}
%
\title{Space-Ground Coherent Optical Links: Ground Receiver Performance With Adaptive Optics and Digital Phase-Locked Loop}
%
%
%


\author{Laurie~Paillier, Rapha\"{e}l~Le~Bidan,~\IEEEmembership{Member,~IEEE,} Jean-Marc~Conan, G\'{e}raldine~Artaud, Nicolas~V\'{e}drenne~and~Yves~Jaou\"{e}n
\thanks{L. Paillier, J-M. Conan and N. V\'edrenne are with ONERA, DOTA, Paris Saclay University, F-92320 Ch\^atillon, France e-mail: laurie.paillier@onera.fr}
\thanks{R. Le Bidan is with IMT Atlantique, Lab-STICC, UBL, 29238 Brest, France}
\thanks{G. Artaud is with CNES, 31400 Toulouse, France}
\thanks{Y. Jaou\"en is with LTCI, T\'el\'ecom Paris, Institut Polytechnique de Paris, 75013 Paris, France}
\thanks{Manuscript received Month Day, Year; revised Month Day, Year.}}

%
%

\markboth{Journal of Lightwave Technology,~Vol.~xx, No.~xx, Month~2019}%
{Shell \MakeLowercase{\textit{et al.}}: Bare Demo of IEEEtran.cls for IEEE Journals}
%



\maketitle

\begin{abstract}
In the framework of optical high data rate satellite-to-ground links, we investigate the performance of a coherent receiver which combines an adaptive optics system and a digital carrier synchronization technique. 
We propose a digital carrier synchronization system based on a phase-locked loop and use end-to-end simulations to characterize the impact of the atmospheric turbulence in the case of a representative low Earth orbit satellite-to-ground transmission link. 
The results show that classical digital phase-locked loop techniques can be a reliable solution to accurately compensate for the Doppler frequency shift between the incoming signal and the local oscillator at the ground station, in the presence of a preliminary coarse frequency estimation. 
Adaptive optics compensation enables the loop to lock after a few milliseconds and to accurately track the phase fluctuations. 
The bit-error-rate performance is not degraded by the digital loop but suffers from the residual signal amplitude fluctuations.
\end{abstract}


%
\IEEEpeerreviewmaketitle

\section{Introduction}
%
%
%
%

\IEEEPARstart{I}{n} terrestrial fiber-optic communication networks, coherent optical transmissions are nowadays a key solution to reach high data rates  \cite{winzer2018fiber}. 
The coupling of phase modulation and coherent detection indeed facilitates wavelength division multiplexing (WDM) and offers higher sensitivity as well as the possibility of high-order modulation compared to direct detection. 
Free-space inter-satellite communications links already benefit from the numerous advantages of using optical transmission rather than or in addition to radio-frequency communications, such as lower power consumption or higher bandwidth \cite{kaushal2016optical}, particularly with coherent methods \cite{heine2014european}. 
Furthermore, recent analytical \cite{belmonte2009capacity} \cite{belmonte2016satellite} and experimental \cite{calvo2019optical} \cite{conroy2018demonstration} studies have contributed to demonstrate the high potential of coherent receiver for high data rate satellite-to-ground communications.  
Uplink, downlink and bilateral communications experiments between a ground station and low Earth orbit (LEO) and geostationary (GEO) satellites based on coherent receiver technology have even been established \cite{saucke2016tesat}.

Both satellite-to-ground links and inter-satellite transmissions undergo Doppler effects due to the relative motion between the emitter and the receiver \cite{Shoji2012}\cite{rosenkranz2016receiver}. 
This results in a frequency shift of the incoming signal with respect to the local oscillator (LO) frequency amplified by the frequency instability of both lasers, which disturbs carrier synchronization. 
In this context, optical phase-locked loops (OPLL) naturally appear as a solution to overcome this frequency mismatch and perform carrier recovery in inter-satellite links \cite{ando2011coherent}\cite{schaefer2015numerical}. 

However, the space-ground channel presents specific characteristics that need to be accounted for. 
The propagation through atmospheric turbulence significantly and randomly impairs the amplitude and phase of the optical wave \cite{roddier1999adaptive}. 
As a result the mixing efficiency between the incoming signal and the LO oscillator collapses dramatically which ultimately reduces the down-converted power.
A state-of-the-art solution to mitigate turbulence-induced phase distortions is the use an adaptive optics (AO) system which improves the spatial phase matching between the signals \cite{saucke2016tesat}. 
However the AO correction is not perfect and residual phase aberrations remain. Moreover, the random amplitude variations, called scintillation, also impair the received signal without being compensated by the AO system. To design an effective carrier recovery strategy, a fine modelling of these turbulent effects is thus crucial.
Previous works have studied the performance of ground-space coherent receivers based on statistical models of the turbulence effects \cite{belmonte2009capacity}\cite{valencia2015atmospheric}. 
Their conclusions clearly show the interest in using coherent modulation schemes. Open questions still remain concerning the feasibility of phase locking in the presence of realistic free-space optical (FSO) channel impairments.

In \cite{conroy2018demonstration} the authors describe a digital coherent receiver based on digital signal processing (DSP) techniques developed for fiber-optic networks, with application to a 10.45 km FSO horizontal link. 
Frequency offset compensation is done by performing a Fourier transform of the signal. 
Then the DSP system applies a Mth-power phase estimation to the signal similar to those used in fiber-optic networks.
A similar approach to phase recovery is presented in \cite{valencia2015atmospheric}. 
Neither of these studies however does focus on the performance of the carrier synchronization components. 
An analog solution combining an OPLL and an optical injection loop is presented in \cite{Shoji2012} to deal with Doppler frequency shifts up to 10 GHz for 10 Gb/s binary phase-shift keying (BPSK) transmission. However the study does not take into account the random amplitude fluctuations induced by atmospheric turbulence. 

In this article, we use an accurate numerical modeling of the optical wave undergoing atmospheric turbulence and AO correction to obtain realistic temporal series of the mixing efficiency. 
We then present a digital closed-loop receiver architecture for carrier synchronization combining a digital automatic gain control (AGC) with a digital phase-locked loop (DPLL), and characterize its performance in the presence of realistic turbulence perturbation. 
PLLs have long been used for coherent demodulation of wireless signals. One finds them in particular at the heart of the robust carrier tracking techniques used in global positioning satellite systems (GNSS) \cite{lopez2013survey} for their ability to track both a frequency offset and the phase fluctuations of the signal induced by the channel. 
However, as with any feedback system, the stability domain of the PLL has to be carefully assessed. 
Compared to optical PLL, DPLL have the advantage of flexibility and ease of reconfigurability.
As the initial frequency mismatch in space-ground FSO links may be of the order of several GHz \cite{Shoji2012}, we assume hereafter that the major part of this offset is pre-compensated with a coarse frequency estimator, based for example on ephemerides, which is out of the scope of the present study. Our system then acquires and tracks the residual shift with the objective of maintaining accurate frequency and phase synchronization during the pass of the satellite. 

The overall architecture of a LEO satellite downlink using AO compensation and the description of the parameters are presented in Part II. The design of the digital carrier synchronization system is presented in Part III with a detailed focus on the design of the DPLL. The system performance in the presence of realistic propagation impairments is investigated in Part IV. Conclusion follow in Section V.

\section{Modeling of a coherent LEO-to-ground link}\label{part2}
\subsection{Overall architecture}

\begin{figure}[h]
\centering
   \includegraphics[width=\linewidth]{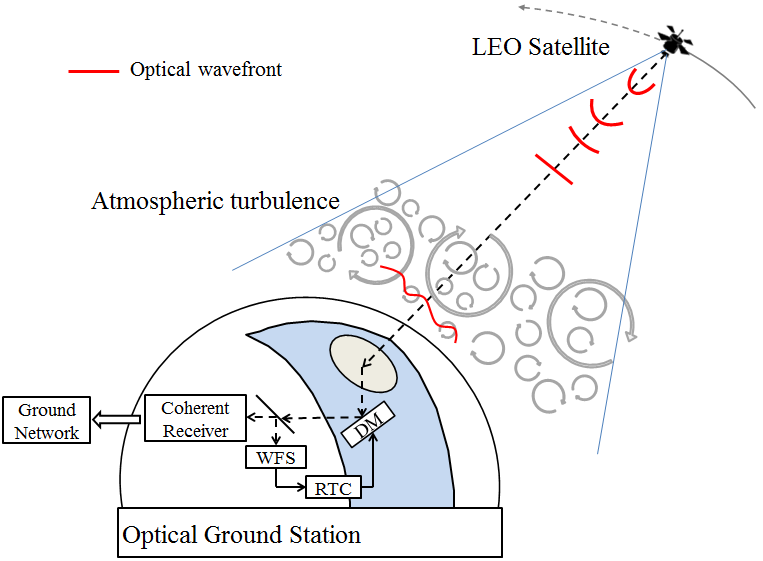}
 \caption{General principle of a satellite-to-ground optical link with adaptive optics correction}
   \label{schemaprincipe}
\end{figure}

\begin{figure*}[!t]
\centering
   \includegraphics[width=\linewidth]{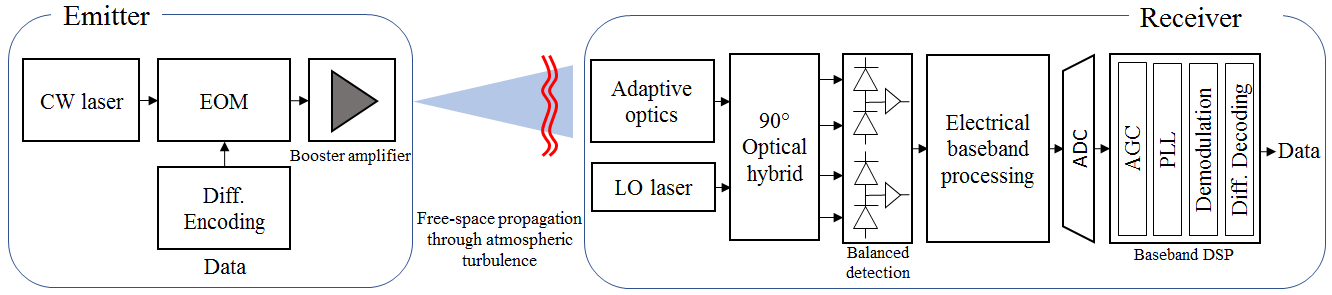}
 \caption{Overall architecture of a LEO-to-ground communication link with coherent detection}
   \label{overallarchi}
\end{figure*}

We consider a LEO-to-ground link having the overall architecture illustrated in Fig.~\ref{schemaprincipe} and Fig.~\ref{overallarchi}. The transmitter at the LEO satellite is located several hundreds of km away from the optical ground station (OGS) receiver. Binary data are first differentially encoded, then transmitted with BPSK by modulating the phase of a continuous wave (CW) laser with the help of an electro-optic modulator (EOM) such as a Mach-Zehnder modulator. Compared to higher order modulation, the BPSK modulator is more convenient to integrate with the additional advantage of requiring less on-board power. The optical signal is then amplified by a booster amplifier and emitted in the direction of the OGS. The impact of the booster amplifier and of the on-board laser phase noise on the optical phase \cite{moller1998novel}\cite{rochat2001new}\cite{ricciardi2013phase}
 will not be discussed in this paper.
During free-space propagation through the atmosphere, the optical wave is severely degraded both in amplitude and in phase by the atmospheric turbulence. 

On the ground, the incoming signal is mixed with the LO.
An AO system is used in order to maximize the mixing efficiency and thus the down-converted coherent power by improving the spatial matching between the incoming and LO fields, respectively. In practice, the AO operates as follows \cite{roddier1999adaptive}.
A wavefront sensor (WFS) returns local information on the incoming wavefront. The WFS data are processed by the real-time computer (RTC) which controls the deformable mirror (DM) and adapts the surface of the latter so as to correct for the wavefront deformation. The resulting optical signal is then coherently detected by an intradyne receiver, sampled by an analog-to-digital converter (ADC), and demodulated by a digital baseband receiver composed of a digital AGC, a DPLL, a symbol detector and a differential decoder. 

\subsection{Channel description and modeling}\label{channeldescription}
Variations of humidity and temperature in the atmosphere induce spatial and temporal random fluctuations of the refractive index of air. 
This phenomenon called atmospheric turbulence disturbs the amplitude and phase of the optical incident beam. The refractive index structure constant $C_n^2(h)$ quantifies the local strength of the turbulence along the line of sight at different altitudes $h$, from the ground level to the maximum altitude $h_{max}$ (here $h_{max}=20~km$).
The $C_n^2(h)$ profile used is derived from the Hufnagel-Valley profile \cite{Valley1980} taken from the recommendation ITU-RP.1621-1. 
This model is parametrized by the value of the $C_n^2$ at the ground level $C_0$ and the high-altitude root-mean-square (RMS) wind speed parameter $v_{RMS}$. A Bufton wind profile \cite{bufton1973comparison} is considered here. The profile is defined by the wind speeds at the ground level $v_G$ and at the tropopause $v_T$.  
The spatial coherence of the optical beam can be characterized by the Fried parameter $r_{0}$.
This parameter gives information on the total turbulence strength along the line of sight. 
The scintillation index $\sigma^2_I$ quantifies the variations of power on the receiver pupil. Besides, the simulation also assumes a given outer scale $L_0$ corresponding to the characteristic size of the larger eddies inducing the energy transfers
in the atmosphere. 
The parameters characterizing the simulated turbulence are given in Table~\ref{table_turb}. In the table, the Fried parameter and the scintillation index are given on the line-of-sight at the wavelength $\lambda=1550~nm$.

\begin{table}[h]
\renewcommand{\arraystretch}{1.3}
\caption{Link Characteristics and Turbulent Parameters for a LEO-to-Ground Scenario}
\label{table_turb}
\centering
\begin{tabular}{c c}
\hline \hline
Parameters & Values\\ 
\hline
Propagation wavelength $\lambda$ & 1550 nm\\ 
$C_0$ & $10^{-13} m^{-2/3}$\\
RMS wind speed parameter $v_{RMS}$ & 20 $m.s^{-1}$\\
Wind speed at the ground level $v_G$ & 10 $m.s^{-1}$\\
Wind speed at the tropopause $v_T$ & 20 $m.s^{-1}$\\
Fried parameter $r_0 $ & 0,039 m\\
Outer scale $L_0$ & 5 m \\
Scintillation index $\sigma^2_{I}$ & 0.684 \\
Elevation & 20 deg\\
Satellite transverse velocity & 6.5 km.$s^{-1}$\\
Rx aperture diameter $D_{RX}$ & 50 cm \\
\hline \hline
\end{tabular}
\end{table}
The turbulent parameters are chosen to simulate a representative channel using data taken from the literature \cite{Kudielka2019}. 
We consider that the communication with the LEO satellite is established at an elevation of 20 degrees to maximize the duration of the data transmission. 
Realistic temporal series of the propagated complex field are obtained using the end-to-end propagation code TURANDOT developed by ONERA in cooperation with CNES \cite{Vedrenne2012}. 
In this method, the turbulent volume is sampled in discrete layers (here 35 layers) and a split-step algorithm performs Fresnel propagation between each discrete phase screen. 
The field entering the atmosphere at the altitude $h_{max}$ is approximated by a plane wave considering the very large propagation distance from the satellite.

The complex field received within the telescope pupil plane is then sent to an end-to-end adaptive optics simulation. We assume that the correction modes are the Zernike polynomials (here up to mode 91, that is 12 radial orders) \cite{Noll:76}. The AO simulation models a closed loop at a given sampling frequency (here 5kHz) with a 2 frame loop delay. The chosen AO design parameters are inspired from \cite{Vedrenne2016} and lead to an average flux penalty of -4.5 dB in the turbulent conditions described above. The WFS noise is neglected since requirements imposed by the data transmission lead to a high flux regime from the WFS point of view.  

\begin{figure*}[h]
 \centering
\subfloat[]{\includegraphics[width=0.20\linewidth]{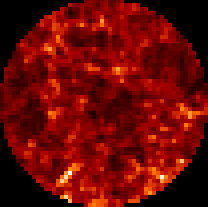}
\label{scinti}}
~
\subfloat[]{\includegraphics[width=0.20\linewidth]{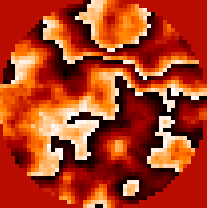}
\label{phaseturb}}
~
\subfloat[]{\includegraphics[width=0.20\linewidth]{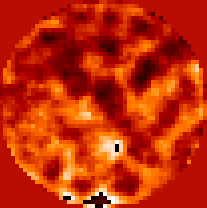}
\label{phaseoa}}
~
\subfloat[]{\includegraphics[width=0.20\linewidth]{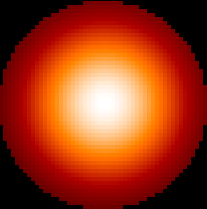}
\label{modefibre}}
\caption{a. Pupil amplitude map due to scintillation b. Turbulent phase in the pupil c. Residual phase after AO correction d. Amplitude of a Gaussian mode at $w_0=\frac{D}{2.2}$. The phase maps are illustrated modulo $2\pi$.}
   \label{phasescint}
\end{figure*}

The electromagnetic field's complex amplitude of the incident beam after propagation  through turbulence and correction with AO is characterized by : 
\begin{equation}\label{ERX}
    E_{RX}(r,t)=A_{TX}exp(\chi(r,t)+i(\phi_{res}(r,t))
\end{equation}
The factor $A_{TX}$ is a function of the emitted power. $\chi$ represents the log-amplitude fluctuation induced by scintillation and $r$ is the transverse two-dimensional spatial coordinate. In the remainder of this section and for the ease of exposition, the modulated phase and the frequency mismatch are for now discarded when defining the complex signal amplitudes. The residual phase $\phi_{res}$  corresponds to the residual phase  perturbation due to the atmospheric turbulence $\phi_{tur}$ compensated by the AO phase correction $\varphi_{AO}$:
\begin{equation}\label{phires}
    \phi_{res}(r,t)=\varphi_{tur}(r,t)-\varphi_{AO}(r,t)
\end{equation}
The LO is represented by a Gaussian mode in the aperture plane \cite{shaklan1988coupling} : 
\begin{equation}\label{ELO}
    E_{LO}(r,t)=A_{LO}exp(-r^2/w_0^2)
\end{equation}
where $A_{LO}$ is the constant amplitude of the LO and $w_0$ is the mode radius. The mode radius is set to $w_0=\frac{D}{2.2}$ to maximize the efficiency while neglecting the central obscuration influence \cite{klein1974optical}. The LO power is denoted by $P_{LO}=\frac{|A_{LO}|^2}{2}$. 
Bi-dimensional scintillation and phase maps are illustrated in Fig.~\ref{scinti} and \ref{phaseturb}. Turbulent phase residuals after AO correction is illustrated on Fig.~\ref{phaseoa}. Fig.~\ref{modefibre} shows the Gaussian distribution of the LO mode projected in the aperture plane.
The mixing efficiency between the incoming signal and the Gaussian mode of the LO is affected by the residual turbulence effect after the AO compensation.
The instantaneous mixing efficiency can be derived from the complex coupling defined for instance in \cite{Winick:86}.
The complex coupling is an overlap integral expressed in the aperture plane, and reads:

\begin{equation}
    \mathcal{C}(t) = \int_P E_{LO}^*(r,t)E_{RX}(r,t)dr
\end{equation}
where the pupil transmittance $P$ is defined by
\begin{equation}
P(r) = \left\{
    \begin{array}{ll}
        1 & if~0 \leqslant 2|r|\leqslant D_{RX} \\
        0 & otherwise 
    \end{array}
\right.
\end{equation}

The instantaneous coupling efficiency $\rho(t)$ and the phase noise $\phi(t)$ coming from the propagation are denoted as:
\begin{equation}\label{rho}
        \rho(t)=|\mathcal{C}(t)|^2  
\end{equation}
\begin{equation}\label{phinoise}
    \phi(t)=arg(\mathcal{C}(t))
\end{equation}

\subsection{Impact of the turbulence on the complex coupling}
The simulation tool provides correlated time series of both the coupling efficiency and the phase noise in the considered turbulent conditions. 
The 2-seconds duration time series presented in this paragraph are those that will be used later in the end-to-end system simulation and performance evaluation described in part~\ref{partresult}.

\begin{figure}[h]
\centering
   \includegraphics[width=\linewidth]{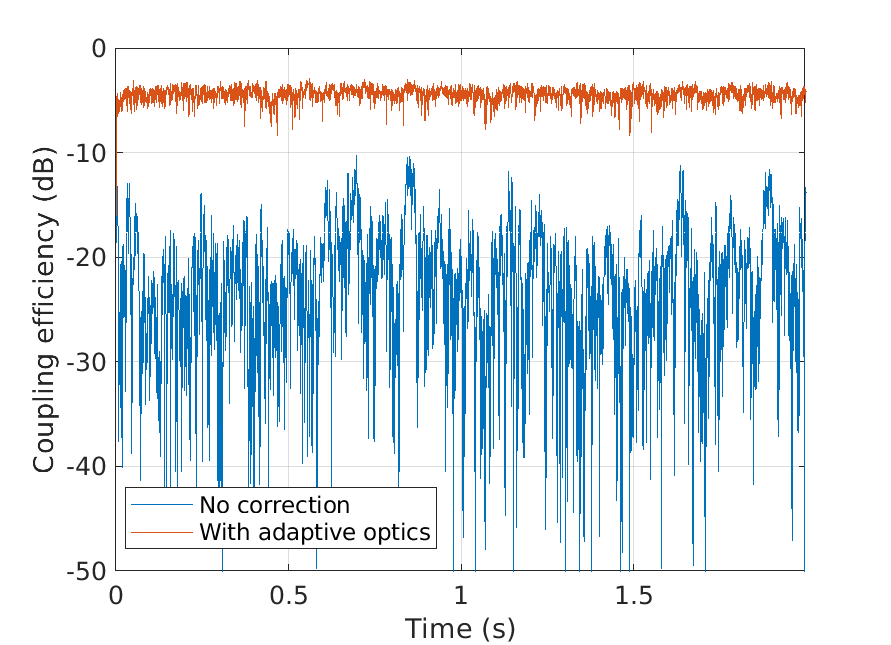}
 \caption{Downlink 2s duration time series of coupling efficiency. Comparison between a time series without correction and with an AO correction of 12 radial orders at the frequency of 5 kHz.}
   \label{tab_abs}
\end{figure}

\begin{figure}[h]
\centering
\subfloat[]{\includegraphics[width=0.5\linewidth]{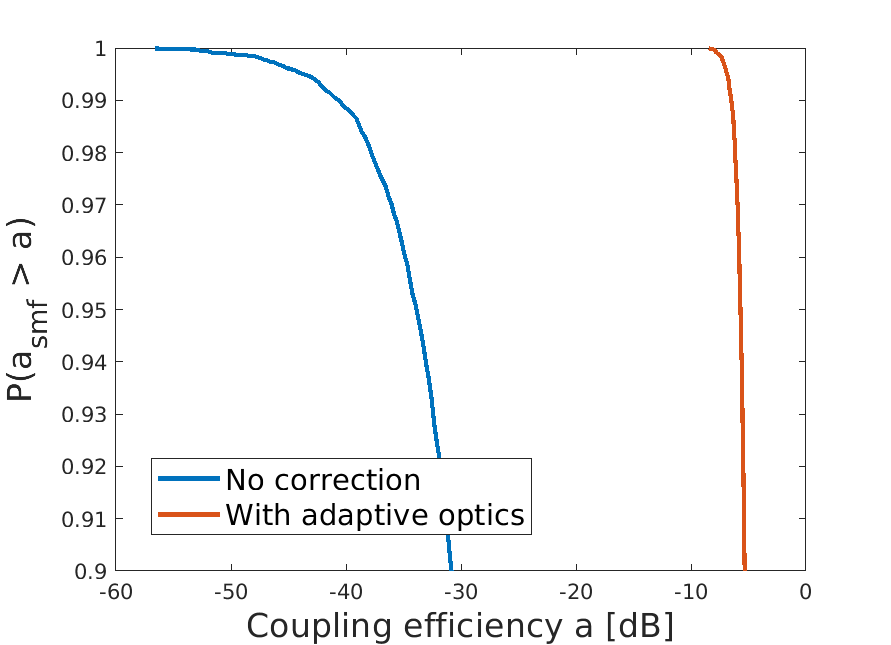} 
\label{Cumul}}
~
\subfloat[]{\includegraphics[width=0.5\linewidth]{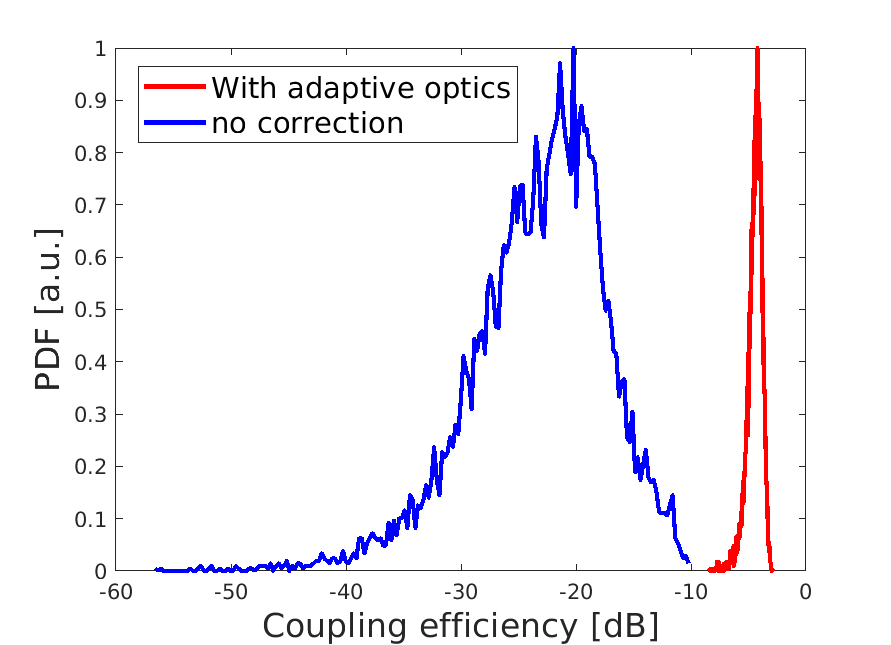}
\label{Hist}}

\caption{Statistics of the coupled efficiency. a. Cumulative probabilities with and without AO correction b. Histogram of simulated coupling efficiency with and without AO correction (arbitrary unit).}
\label{densfunc}
\end{figure}

Fig.~\ref{tab_abs} presents the coupling efficiency $\rho$ of the optical flux with and without AO compensation using the parameters described in the section~\ref{channeldescription}. 
Fig.~\ref{tab_abs} shows that the average flux penalty due to the turbulence is reduced from -23 dB to -4.5 dB thanks to the AO correction. 
Besides the cumulative density function and the probability density function (PDF) presented in Fig.~\ref{densfunc} provide a quantification of the fading probability of occurrence and also prove that the system reduces the amplitude of the signal fluctuations. 

\begin{figure}[h]
\centering
   \includegraphics[width=\linewidth]{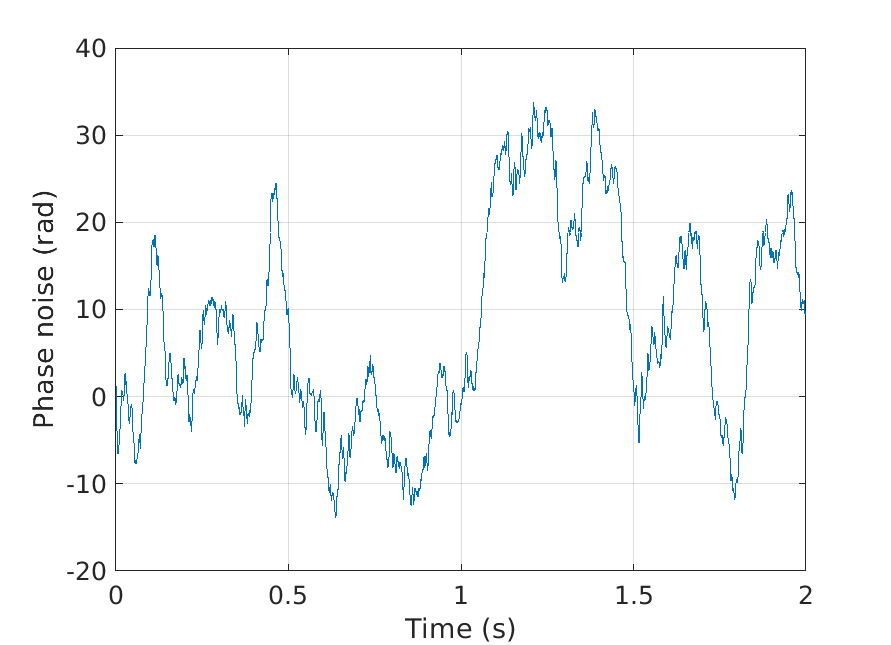}
 \caption{Downlink 2s duration time series of phase noise with an AO correction of 12 radial orders.}
   \label{phase_noise}
\end{figure}

Fig.~\ref{phase_noise} presents the turbulent phase noise $\phi$.  
Most of the residual phase noise can be attributed to the turbulent piston mode which is not corrected by traditional adaptive optics systems \cite{Robert2016}.
As shown on Fig.~\ref{phase_noise} the coherence time of the fluctuations are of the order of 1 ms. It is slower than the phase noise fluctuations of the laser sources currently in use in terrestrial fiber networks. 
This phase noise is also much slower than the symbol rate of 10 Gbaud. We still quantify the effect of the turbulent phase noise in section~\ref{IVC}. 

\section{Coherent detection and baseband digital signal processing design}
The signal after coherent detection and digitization is described in part~\ref{partcoh}. Then, the overall digital carrier synchronization architecture is presented in part~\ref{partdsp} and is followed by a description of the digital AGC loop (see part~\ref{partagc}) and PLL (see part~\ref{partpll}). The design and  characterization of the DPLL are presented in part~\ref{partpll} and \ref{partvalidpll}, respectively, for a constant amplitude signal with BPSK modulation. Results in the presence of turbulent amplitude fluctuations and phase noise will be presented in part~\ref{partresult}.   

\subsection{Coherent intradyne detection and digitization}\label{partcoh}
During the pass of the satellite, the data signal experiences a frequency shift because of the Doppler effect and because of the natural frequency drift of the satellite laser source due to temperature variations for instance. 
The Doppler shift due to the relative speed of a LEO satellite with respect to the ground has been estimated in \cite{Shoji2012} to cover a total range of approximately 9 GHz (from -4.5 GHz to 4.5 GHz). 
This frequency shift is critical for the carrier synchronization between the incoming signal and the local oscillator. 
In the following, this intermediate frequency is denoted $\Delta f = \frac{\omega_{RX}-\omega_{LO}}{2\pi}=\frac{\Delta\omega_0}{2\pi}$. 
We consider that the major part of the frequency mismatch induced by the Doppler effect is pre-compensated by a preliminary coarse frequency estimation stage based for instance on the knowledge of the satellite trajectories. Similarly to inter-satellites communications, a residual frequency shift still remains, due among other things to the instability of the frequency of the emitter laser and of the LO.  
We further assume that this residual shift is of the same order as those encountered in inter-satellite communication \cite{ando2011coherent} \cite{schaefer2016coherent}. Hereafter a maximum constant frequency offset of 100 MHz is considered.  

After AO correction, the optical signal is sent to the coherent receiver and mixed with the LO signal for frequency down-conversion. 
After the balanced detection the signal is sampled and digitized in the form of the real part $s_{RX,I}$ and imaginary part $s_{RX,Q}$ of the down-converted complex amplitude, respectively, where
\begin{equation}\label{rxI}
    s_{RX,I}(k)\propto \sqrt{\rho(k)} \cos(\Delta\omega kT+\varphi_m(k)+\phi(k))
\end{equation}
\begin{equation}\label{rxQ}
    s_{RX,Q}(k)\propto \sqrt{\rho(k)} \sin(\Delta\omega kT+\varphi_m(k)+\phi(k))
\end{equation}
Eq.~\eqref{rxI} and Eq.~\eqref{rxQ} are relative to the discrete-time sample collected in the time-interval $[kT;(k+1)T[$ where $T$ denotes the symbol time. Ideal symbol synchronization is assumed. BPSK modulation is included here in the modulated phase term  $\varphi_m(k) \in \{0,\pi\}$. These equations can be derived from Eq.~\eqref{rho} and Eq.~\eqref{phinoise} by further accounting for the sampling process, frequency mismatch and BPSK phase modulation. Note that $\sqrt{\rho(k)}$ corresponds to the usual $\sqrt{P_{RX}P_{LO}}$ term related to the received power and LO power. Hereafter the LO power is assumed to be larger than the signal power so that the dominant noise source is shot noise, modeled as an additive white Gaussian noise process \cite{barry1990performance}. 

In the following we consider that the coupling efficiency $\rho$ and phase noise $\phi$ induced by the propagation through the atmosphere, described in Fig.~\ref{tab_abs} and Fig.~\ref{phase_noise}, remain constant during a symbol time. 

\subsection{Digital receiver architecture}\label{partdsp}
\begin{figure}[h]
\centering
\includegraphics[width=\linewidth]{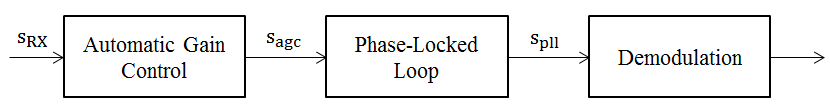}
 \caption{Digital receiver architecture under consideration.}
  \label{archidsp}
\end{figure}

In this part, we describe a baseband digital receiver for data demodulation and detection which operates at the symbol rate and consists of a PLL assisted by an AGC loop, followed by the symbol detector. Fig.~\ref{archidsp} depicts the overall architecture.

First the signal coming from the ADC is sent to an AGC loop described in part~\ref{partagc}. The role of the AGC system is to maintain the received signal power at a constant target value, thereby reinforcing the robustness of the PLL against amplitude fluctuations.

The signal is then sent to the DPLL which first locks onto the frequency shift, then tracks the phase and frequency small variations of the signal. 
The design of the DPLL is described and discussed in part~\ref{partpll}. A BPSK symbol detector and a differential decoder follow in order to recover the transmitted bits from the symbol-rate discrete-time sample at the DPLL output. These two operations are standard practice in digital communication systems and will not be further detailed here. 

\subsection{Digital Automatic Gain Control}\label{partagc}
\begin{figure}[h]
\centering
\includegraphics[width=\linewidth]{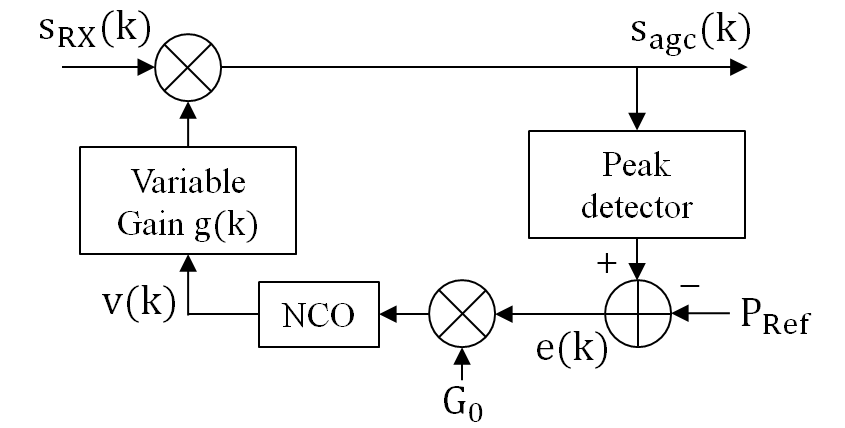}
 \caption{Block diagram of the digital automatic gain control loop.}
  \label{agc_diag}
\end{figure}

The role of the digital AGC is to maintain the signal power at a constant value. The block diagram of the  digital AGC architecture retained in this work is presented in Fig.~\ref{agc_diag}. 

The signal power is first compared to a target reference power $P_{Ref}$ which controls the desired AGC output amplitude, to form the error signal  
\begin{equation}
    e(k)=|s_{agc}(k)|^2-P_{Ref}
\end{equation}
Hereafter a unit power $P_{Ref}=1$ is chosen as reference. The error signal $e(k)$ is then multiplied by a constant gain factor $G_0$ and passed through a numerically controlled oscillator (NCO), here in the form of a first-order integrator with transfer function $1/(z-1)$. The AGC gain was set to $G_0=0.1$ as a trade-off between stability and convergence speed of the loop. Based on the control signal delivered by the NCO, the AGC adjusts the amplitude of the incoming signal $s_{RX}$ through the variable gain function $g(k)$. An exponential characteristic was chosen to ensure that the acquisition time only depends on the AGC parameters and not on the dynamics input signal \cite{Ohlson1974} :
\begin{equation}
    g(k)=exp\left(-\frac{v(k)}{2}\right) 
\end{equation}
$v(k)$ corresponding to the output signal of the NCO. 

\subsection{Digital Phase locked loop}\label{partpll}
\begin{figure}[h]
\centering
\includegraphics[width=\linewidth]{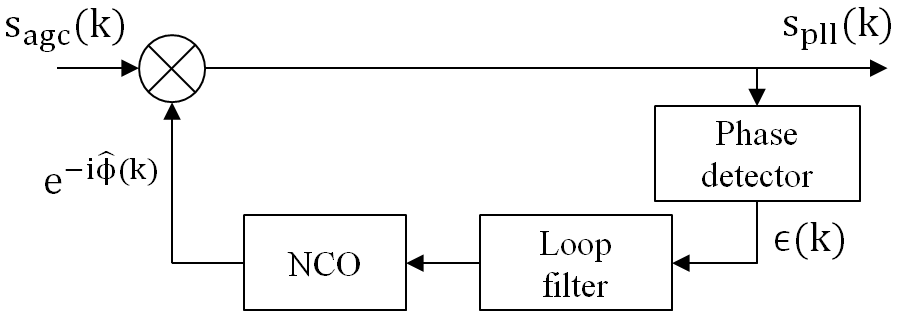}
 \caption{Block diagram of the digital phase locked loop.}
  \label{pll_diag}
\end{figure}

The purpose of the DPLL is to synchronize accurately the phase and frequency of the local oscillator with the received signal so as to minimize the residual phase error between the two signals. Fig.~\ref{pll_diag} shows the block diagram of the PLL under consideration. The loop consists of 3 main components: a phase detector, a loop filter and an NCO. 

In the remainder of this section, the incoming signal amplitude is assumed to be constant. Performance in the presence of turbulence, amplitude fluctuations and phase noise will be investigated later in Part~\ref{partresult}. 
From standard detection theory results, the optimal MAP phase detector for estimating an unknown carrier phase offset affecting a BPSK modulated signal without prior knowledge on the transmitted data is \cite{simon2006carrier}:
\begin{equation}
    \epsilon(k)=s_{pll,Q}(k)\,\tanh{(s_{pll,I}(k))}
\end{equation}
where $s_{pll,I}$ and $s_{pll,Q}$ are respectively the real and the imaginary parts of the signal. 
At low signal-to-noise ratio (SNR) this error signal is well-approximated by \cite{Simon1979} :
\begin{equation}
    \epsilon(k)=s_{pll,I}(k)\,s_{pll,Q}(k)
\end{equation}
The corresponding phase detector has a semi-sinusoidal response in the absence of noise:
\begin{equation}\label{epsk}
    \epsilon(k)=\frac{K_d}{2}\,\sin(2\varphi(k))
\end{equation}
where $K_d$ is the average signal power and $\varphi(k)=\Delta\omega\,k$ is the discrete-time phase offset to be estimated. Note that the estimation of the phase error depends on the amplitude of the incoming signal through the factor $K_d$. Knowledge of the latter is required in the PLL design.  

The phase error signal delivered by the phase detector is smoothed by a second-order loop filter having gains $K_1$ and $K_2$ and discrete-time transfer function:
\begin{equation}
    F(z)=K_1 \left ( 1+\frac{K_2}{z-1} \right)
\end{equation}
The result is sent to the NCO which is implemented as a first-order integrator with transfer function $K_0/(z-1)$ and unit gain $K_0=1$. The phase error estimate is finally removed from the incoming signal phase of the next sample, thereby closing the loop.

Designing the PLL requires choosing the gain coefficients $K_d$, $K_1$ and $K_2$. To this end and following common practice in the field, we chose to work with three intermediate variables having more physical relevance, namely the damping factor $\xi$, the loop bandwidth $B_L$ and the natural frequency $\omega_n$. Those parameters are defined in the literature \cite{gardner2005} for analog loops from the linearized system transfer function assuming that the phase error is small. By applying this method to the digital case, Eq.~\eqref{epsk} is linearized as $\epsilon(k)=K_d\varphi(k)$, and we obtain the following discrete-time equivalent formulas for our three parameters of interest
\begin{equation}
    B_LT=\frac{1}{4}(K+K_2)
\end{equation}  
\begin{equation}
    \xi=\frac{1}{2}\sqrt{\frac{K}{K_2}}
\end{equation}
\begin{equation}\label{eqwn}    
    \omega_nT=\sqrt{KK_2}
\end{equation}
where $K=K_d K_1 K_0$ denotes the global loop gain. It is noted in \cite{gardner2005} that the analog formulas are good approximations for the DPLL parameters as long as $\omega_n$ is small compared to the sampling rate, here equal to the symbol rate 10 GBaud. For our design and based on Eq.~\eqref{eqwn}, we obtain that $\omega_n=9.3~MHz$. This condition is thus fulfilled. 
The design parameters adopted here have the additional advantage that they can be easily related to the pull-in time which measures the time required by the loop to establish the lock. 
Adopting the same approach as before, based on the pull-in time definition for analog loops \cite{gardner2005}, we obtain the following formula for a signal without amplitude fluctuation and a target frequency offset $\Delta\omega$:
\begin{equation}\label{tp}
 T_p=\frac{2\Delta\omega^2}{\xi\omega_n^3}
 \end{equation}
It is common practice in the field to give the damping factor the value  $\xi=\frac{1}{\sqrt{2}}$ as a trade-off between stability of the loop and speed of convergence \cite{gardner2005}. 
The duration of a LEO satellite pass is only of a few minutes so the pull-in time of the loop needs to be much shorter to maximize the duration of the data transmission. 
The normalized loop bandwidth $B_LT$ characterizes the sensitivity of the loop to cycle slips due to noise but also directly affects the acquisition range as well as the convergence speed of the loop. Consequently and given our target frequency mismatch of 100 MHz, we set $B_LT=0.0005$ to ensure that such an offset falls within the loop acquisition capabilities with a reasonably small pull-in time compared to the satellite pass duration and to minimize at the same time the probability of cycle slips occurrence during the pass. 
Knowledge of the damping factor $\xi$ and of the normalized loop bandwidth $B_LT$ is enough to set the remaining loop gain factors $K_1$ and $K_2$ provided the average signal power $K_d$ is known. Our design assumes unit incoming power $K_d=1$ (an hypothesis reinforced by the AGC). The main parameters of the resulting DPLL are summarized in Table~\ref{table_pll}.

\begin{table}[h]
\renewcommand{\arraystretch}{1.3}
\caption{Phase-Locked Loop Main Parameters}
\label{table_pll}
\centering
\begin{tabular}{c c}
\hline \hline
Parameters & Values \\
\hline
$T$ & 0.1 ns\\
Damping factor & $\frac{1}{\sqrt{2}}$\\
Loop bandwidth $B_L$ & 5 MHz\\
$K_1$ & $1.3.10^{-3}$\\
$K_2$ & $6.7.10^{-4}$ \\
Loop frequency & 10 GHz \\
\hline \hline
\end{tabular}
\end{table}

\subsection{PLL characterization without turbulence} \label{partvalidpll}
\begin{figure}[h]
\centering
   \includegraphics[width=\linewidth]{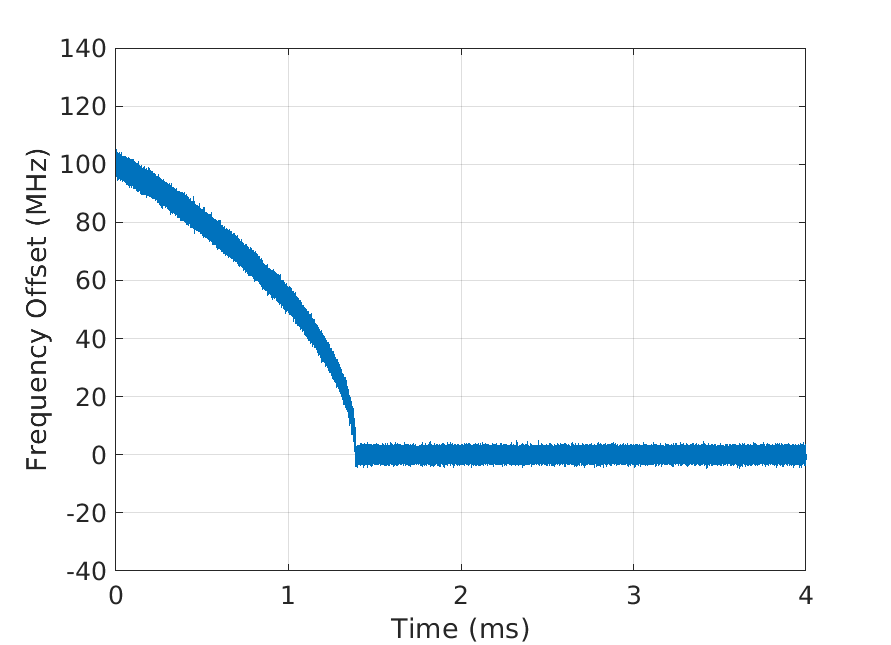}
 \caption{Frequency offset evolution during lock-in at SNR $E_s/N_0=8$ dB}
   \label{validconv}
\end{figure}

Fig.~\ref{validconv} shows the acquisition phase of the presented system. Here, the loop locks after 1.4 ms for an initial frequency offset $\Delta f$= 100 MHz. The DPLL is thus able to acquire the target frequency shift in a reasonable time compared to the duration of a satellite pass. 
Furthermore and as expected, the measured acquisition time is in accordance with the value obtained from Eq.~\eqref{tp}, which predicts $T_p \approx 1.4$ ms. 

\begin{figure}[h]
\centering
   \includegraphics[width=\linewidth]{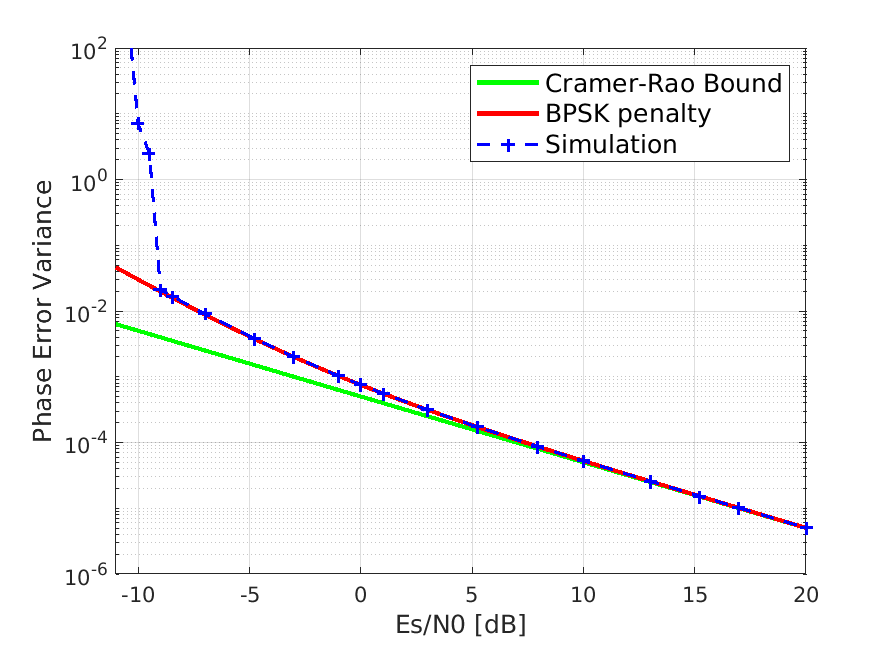}
 \caption{Variance of the phase error at the DPLL output as a function of the SNR and comparison with the theoretical bounds}
   \label{validdpll}
\end{figure}

In the presence of noise, the error signal $\epsilon(k)$ consists of the signal term $\frac{K_d}{2}\,\sin(2\varphi(k))$ plus terms involving the products signal $\times$ noise and noise $\times$ noise. As for any other estimator, the loop performance in the presence of noise is lower-bounded by the Cramer-Rao bound (CRB) which, for an analog loop operating on a constant amplitude sinusoidal signal without phase modulation, reads
\begin{equation}
    \sigma_{CRB}^2=\frac{B_LT}{\frac{E_s}{N_0}}
\end{equation}
In addition, the variance of a linear-operating PLL is known to meet this bound \cite{gardner2005}.
In the case of a BPSK modulated signal, the modulation induces a penalty on the previous bound, the so-called squaring-loss penalty \cite{simon2006carrier}. The variance of the error signal at high-enough SNR $\frac{E_s}{N_0}$ is then well-approximated by 
\begin{equation}\label{sigma2phi}
    \sigma_{BPSK}^2=\frac{B_LT}{\frac{E_S}{N_0}}\times\frac{2\frac{E_S}{N_0}}{2\frac{E_S}{N_0}+1}
\end{equation}
To validate the behavior of the DPLL in the presence of additive white gaussian noise (AWGN), we have plotted in Fig.~\ref{validdpll} the phase error variance as a function of SNR in steady-state performance (transients occurring during the acquisition period have been excluded from the calculation). We first note the existence of a minimum SNR value, here around -9 dB, under which the DPLL is unable to lock and becomes unstable. Above this critical SNR value, the measured phase error variance quickly converges towards the theoretical bound given by Eq. \eqref{sigma2phi}. The latter reaches the CRB as the SNR further increases and the squaring-loss becomes asymptotically negligible. This validates our design. At SNR values lower than the critical SNR, the previous design assumptions and formulas no longer apply since the DPLL operates in a highly non-linear regime and proves to be unable to reach a stable state. However we do not expect the system to operate in such conditions. 

\section{System performance with adaptive optics}\label{partresult}

Having validated the proposed digital carrier recovery system in the ideal case of a BPSK modulated signal impaired by AWGN only, we now investigate its robustness in more realistic propagation conditions based on the FSO link model described in Section~\ref{part2}, more precisely on the simulated time series obtained after AO correction and presented in Fig.~\ref{tab_abs} and \ref{phase_noise}, respectively. First the role of the AGC loop in such an uncertain environment is highlighted. Then we characterize the convergence and tracking performance of the DPLL in the presence of amplitude fluctuations and turbulent phase noise. We finally evaluate the numerically estimated bit-error rate (BER) performance of the overall transmission system. 

\subsection{Fluctuations mitigation with automatic gain control}

\begin{figure}[!h]
\centering
\subfloat[]{\includegraphics[width=0.5\linewidth]{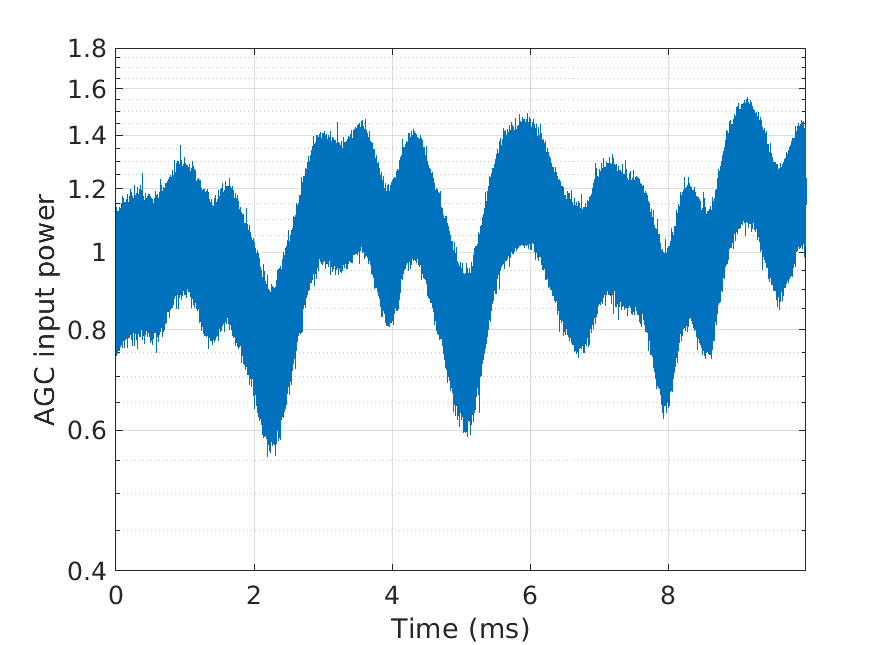} 
\label{fig_first_case}}
~
\subfloat[]{\includegraphics[width=0.5\linewidth]{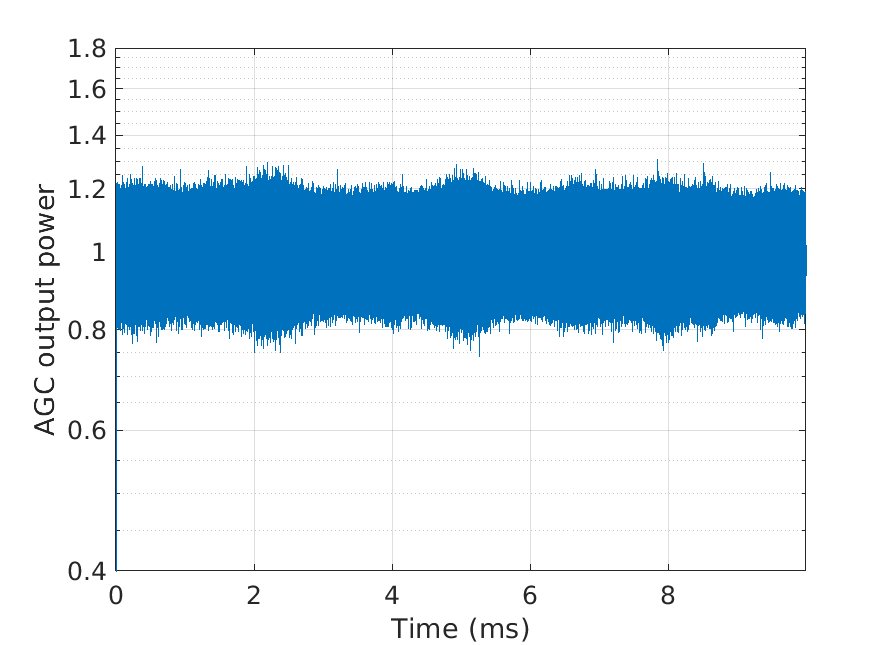}%
\label{fig_second_case}}

\caption{10ms duration time series of the AGC (a) input and (b) output power at an average SNR of 30dB (value chosen for illustration purpose only).}
\label{power_agc}
\end{figure}

The main purpose of the digital AGC is to compensate for the slow residual amplitude fluctuations caused by atmospheric turbulence that remain after AO correction. A unit target reference power $P_{Ref}=1$ is used in the loop so as to match the assumption of unit signal gain $K_d=1$ used in our DPLL design (see Section~\ref{partpll}.D), thereby ensuring that the loop will run in nominal operating conditions. Fig.~\ref{power_agc} shows the impact of the AGC on the received signal power variations. Note that the multiplicative gain of the AGC loop impacts equally the signal and the noise. In particular, in deep fades, both the signal and the noise are amplified. 

\subsection{Digital frequency shift compensation}
\begin{figure}[h]
\centering
   \includegraphics[width=\linewidth]{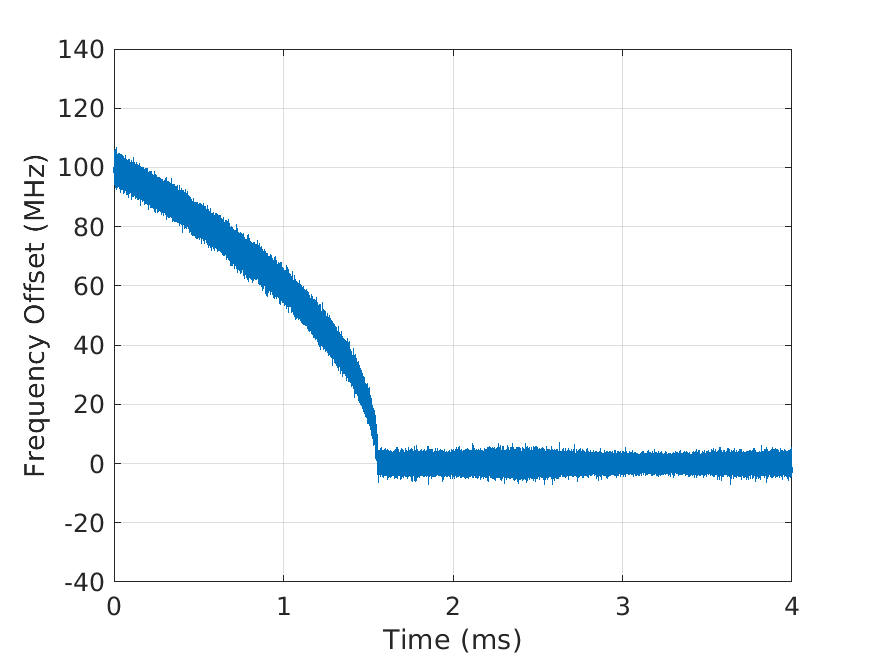}
 \caption{Frequency error during the lock-in process. Temporal series with a mean SNR of 8 dB}
   \label{acquisition}
\end{figure}

After the AGC loop the signal is sent to the PLL. Fig.~\ref{acquisition} presents the acquisition step of the PLL at an average SNR of 8dB. 
The initial frequency offset between the incoming signal and the LO was again set to 100 MHz. 
It is interesting to note that despite the residual amplitude fluctuations, the theoretical prediction  still applies since the DPLL requires again 1.4 ms only to converge. This confirms that with proper prior AO correction and AGC compensation, the proposed digital carrier synchronization solution can maintain very short lock-in time compared to the duration of the satellite pass even in turbulent conditions. 

\begin{figure}[!h]
\centering
   \includegraphics[width=\linewidth]{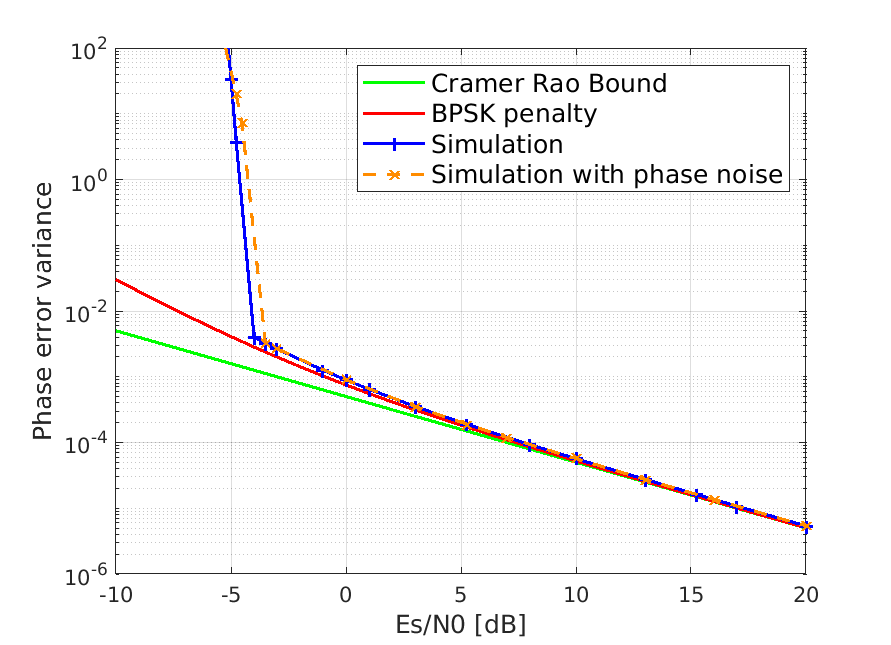}
 \caption{Phase error variance as a function of the mean SNR of the 2s-duration time series with and without turbulent phase noise after the DPLL compared to the theoretical expressions.}
   \label{msevssnr}
\end{figure}

Fig.~\ref{msevssnr} shows the evolution of the variance of the phase error at the DPLL output (excluding the acquisition period) as a function of the average SNR on the overall time series. 
The results are compared to the theoretical expression given in Eq.~\eqref{sigma2phi} and derived for a constant amplitude signal and an analog loop. 
The results closely match the bound prediction at average SNRs of 8 dB and higher. This proves that the DPLL is able to maintain the lock and accurately track the residual phase fluctuations even in the presence of turbulence. Interestingly, in the present case, the main impact of fading is to increase the minimal critical SNR value under which the DPLL becomes unstable by approximately 5 dB compared to the case without amplitude fluctuation, as can be seen by comparing Fig.~\ref{msevssnr} and Fig.~\ref{validdpll}.

\subsection{Impact of turbulent phase noise with adaptive optics correction}\label{IVC}
To confirm that the turbulent phase noise $\phi$ has no impact on the proposed carrier synchronization system performance, the same simulation is done with an initial frequency offset $\Delta f=100~MHz$ and the turbulent phase noise shown in Fig.~\ref{phase_noise}. The variance of the residual phase error measured at the DPLL output has been superimposed on the other results in Fig.~\ref{msevssnr}. Negligible differences are observed with respect to the performance without turbulent phase noise. 

\subsection{BER performance}
\begin{figure}[!h]
\centering
   \includegraphics[width=\linewidth]{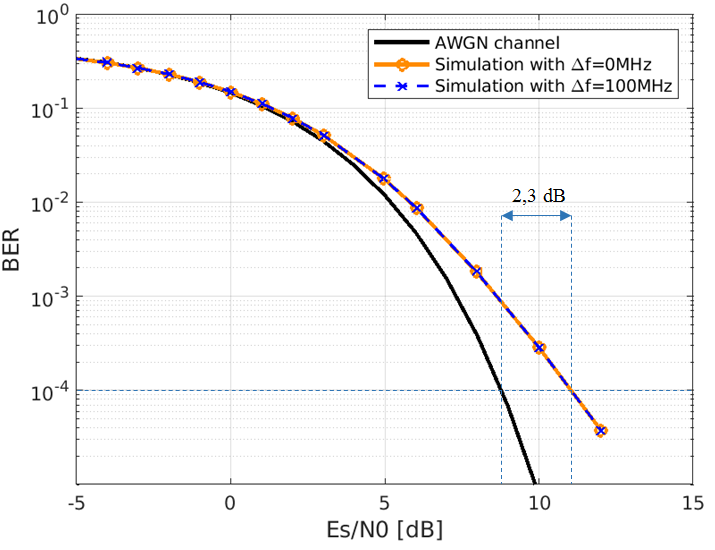}
 \caption{BER performance as a function of the average SNR}
   \label{BER}
\end{figure}

Once the amplitude variations and the frequency offset have been corrected, BPSK symbol detection is performed, followed by differential decoding in order to remove the phase ambiguity which may subsist when the PLL locks onto an incorrect phase due to the rotational symmetry of the PSK constellation.  
Fig.~\ref{BER} shows the BER performance of the simulated system evaluated on the 2s-duration time series presented in Fig.~\ref{tab_abs} as a function of the average SNR. The BER is calculated once convergence is acquired. We first note that the BER performance is virtually the same with and without frequency offset (set to 100 MHz in the simulation). This demonstrates the capability of the proposed digital carrier recovery solution to correct the target frequency offset without BER penalty. For reference purpose, the theoretical BER performance for differentially-encoded BPSK with coherent detection and differential decoding over AWGN is also shown. By comparing the two curves, the fadings caused by atmospheric turbulence are seen to result in a 2.3~dB power penalty at a BER=$10^{-4}$.

\section{Conclusion}
Coherent receivers are gradually emerging as a key enabler for high-data-rate communications not only in fiber-optics networks but also in free-space optical communications. In this paper, we have investigated end-to-end link modeling and receiver architecture design for a coherent 10 Gb/s BPSK LEO satellite-to-ground FSO transmission in the presence of realistic atmospheric turbulence. At the ground station, AO correction is used to improve the mixing efficiency of the incoming signal with the LO and achieve an average flux penalty of -4.5 dB. The impact of the propagation and AO correction on the phase noise has also been characterized. 
After intradyne detection, a digital receiver architecture has been proposed that combines an AGC with a PLL in order to recover from Doppler frequency offsets and track residual phase variations in the presence of atmospheric turbulence. The overall performance of this system has been characterized based on realistic time series of a representative LEO-to-ground link with AO correction obtained from end-to-end simulations. The AO correction enables the system to operate and limits the effect of turbulent amplitude fluctuations of the coupled flux. The digital receiver is able to acquire and track frequency offsets up to 100 MHz in a few ms, a reasonable amount of time compared to the pass duration of a LEO satellite, and without BER penalty with respect to the ideal case of a LO perfectly synchronized in phase and frequency with the received signal. We however show that amplitude fluctuations degrade the BER performance and we quantify this effect.
We also demonstrate that the turbulent phase noise has negligible impact on the carrier synchronization process. 
In this paper, the DPLL was shown to be a simple and robust solution to the problem of Doppler frequency offset correction and phase tracking in coherent satellite-to-ground FSO links. Other solutions may however be considered. In particular, the performance and robustness of open-loop carrier synchronization algorithms such as those used in fiber-optics networks should also be investigated in similar conditions. Also, the DPLL was specifically designed for BPSK modulation in this work. The same methodology can however be extended in a straightforward manner to higher-modulation formats by modifying accordingly the phase detector function within the loop. Finally, all the present work assumed ideal timing recovery. In practice the latter will also be impaired by the Doppler effect and by the signal fadings prompting the need for robust digital symbol synchronization algorithms.


%



\section*{Acknowledgment}
This work was conducted in the framework of a PhD thesis co-funded by ONERA and CNES.

\ifCLASSOPTIONcaptionsoff
  \newpage
\fi



\bibliographystyle{IEEEtran}
%




\bibliography{bibliography} 

%
\begin{IEEEbiographynophoto}{Laurie Paillier}
received her Master of Science in Optics Engineering at the T\'el\'ecom Saint-Etienne Engineering School, Saint-Etienne, France, in 2017. She is currently pursuing her PhD in Optical Telecommunication at ONERA, the French Aerospace Lab, and the Institut Polytechnique de Paris, France, in collaboration with T\'el\'ecom Paris, IMT-Atlantique and CNES. Her research focuses on optical coherent links from space to ground using adaptive optics systems. 
\end{IEEEbiographynophoto}
\begin{IEEEbiographynophoto}{Raphael Le Bidan}
(M'03) received the Eng. Degree in Telecommunications and the M. Sc. Degree in Electrical Eng. from the Institut National des Sciences Appliquées (INSA), Rennes, France, in June 2000, and the Ph. D. degree in Electrical Eng. from the INSA, Rennes, in November 2003. 
Since December 2003, he is working as an associate professor at IMT Atlantique (formerly ENST Bretagne), in the Signal \& Communications dept. Much of his research interests currently focus on forward error codes and digital receiver design for coherent high-throughput fiber and free-space optical transmission systems.
\end{IEEEbiographynophoto}

\begin{IEEEbiographynophoto}{Jean-Marc Conan}
received the Ph.D. degree from Institut d'Optique Graduate School, France, in 1994.
He has been working for more than 20 years in the High Angular Resolution unit of ONERA, the French Aerospace Lab. His major publications focus on:
modeling of turbulence optical effects, optimal control for adaptive optics and wide field tomographic adaptive optics. Originally working in the field of astronomical imaging, he now explores other applications that benefit from this background: impact of turbulence on ground-space optical links for either high capacity telecommunications or high precision frequency transfer.
\end{IEEEbiographynophoto}
\begin{IEEEbiographynophoto}{G\'{e}raldine Artaud}
has been working for CNES since 2006. She is working on space to earth high data rate communications in RF and optics. She is involved in the study of future systems using optical data transmissions for space. She is the responsible at CNES of the DOMINO demonstrator.
\end{IEEEbiographynophoto}
\begin{IEEEbiographynophoto}{Nicolas V\'{e}drenne}
received the Ph.D degree from Nice-Sophia Antipolis University, France, in 2008.
He is the manager of the High Angular Resolution unit of ONERA, the French Aerospace Lab. The research unit has a long term experience in optical wave propagation through the atmosphere and its correction by adaptive optics. The team has developed TURANDOT, the wave optics code dedicated to the simulation of atmospheric turbulence on ground-satellite laser links. It participates to the development of turbulence mitigation techniques, including definition of interleaving and error correcting codes. It is involved in the DOMINO project, CNES demonstrator for laser communications, and leads the FEEDELIO project for ESA aiming at the experimental demonstration of the pre-compensation by adaptive optics for GEO feeder links.
\end{IEEEbiographynophoto}
\begin{IEEEbiographynophoto}{Yves Jaou\"{e}n}
received the Ph.D degree in Physics from Ecole Nationale Sup\'{e}rieure des T\'{e}l\'{e}communications (ENST), Paris, France in 1993, then his HDR (Research management certificate) in 2003. He had joined ENST (now called T\'{e}l\'{e}com Paris) in the Communications and Electronic department in 1982 where he is currently professor. He is lecturing in the domain of electromagnetic fields, optics and optical communications systems. His present researches include high bit rate coherent optical communication systems including digital signal processing aspects, new characterization techniques for advanced photonic devices, high power fiber lasers, fiber optics and remote sensing. He is author or co-author of more 230 papers in journals and communications.
\end{IEEEbiographynophoto}






\end{document}